\documentclass[pra,preprint,superscriptaddress,amssymb]{revtex4-1}
\usepackage{graphicx}% Include figure files
\usepackage{bm}
\usepackage{amsmath,amssymb,fancybox}
\usepackage{color}
\usepackage{theorem}

\theorembodyfont{\normalfont}
\newtheorem{theorem}{Theorem}

\begin{document}
\begin{flushright}
YITP-20-29
\end{flushright}

\title{Information-theoretically-sound
non-interactive
classical verification of quantum computing
with trusted center} 
\author{Tomoyuki Morimae}
\email{tomoyuki.morimae@yukawa.kyoto-u.ac.jp}
\affiliation{Yukawa Institute for Theoretical Physics,
Kyoto University, Japan}
\affiliation{PRESTO, JST, Japan}

\begin{abstract}
The posthoc verification protocol 
[J. F. Fitzsimons, M. Hajdu{\v s}ek, and T. Morimae,
Physical Review Letters {\bf120}, 040501 (2018)] enables 
an information-theoretically-sound
non-interactive verification of quantum computing, 
but the message from the prover to the verifier is quantum
and the verifier has to do single-qubit measurements.
The Mahadev protocol removes these quantum parts,
but the soundness becomes the computational one.
In this paper, we construct
an information-theoretically-sound
non-interactive classical verification protocol for quantum computing 
with a trusted center.
The trusted center sends random BB84 states 
to the prover, and the classical descriptions of these BB84 states
to the verifier.
The messages from the center to the prover and the verifier are
independent of the instance.
By slightly modifying our protocol, we also construct
a non-interactive statistical 
zero-knowledge proof system for QMA with the trusted center.
\end{abstract}
\maketitle  

\section{Introduction}
Whether quantum computing is classically 
verifiable or not is one of the most important open problems
in quantum information~\cite{Gottesman,AharonovVazirani,Andru_review}. 
There have been many partial solutions to the open problem.
These results are categorized into the following six types of
approaches.
\begin{itemize}
\item
Slightly quantum verifier: If the verifier can do some minimum
quantum operations, such as single-qubit generations or measurements,
quantum computing can be 
verified~\cite{FK,Aharonov,HM,MNS,posthoc,Barz,Chiara,Andru,TakeuchiMorimae}.
\item
Multiple provers: If more than two provers who are entangled but not allowed
to communicate with each other are available, a 
classical verifier can verify quantum 
computing~\cite{MattMBQC,Ji,RUV,Grilo,Coladangelo}.
\item
Computational soundness: If the LWE problem~\cite{Regev} is hard for
polynomial-time quantum computing,
quantum computing is classically verifiable
with the soundness against a quantum polynomial-time 
prover~\cite{Mahadev,AndruVidick,Alagic,Yamakawa}.
\item
Sum-check:
BQP is in IP, and therefore
quantum computing is classically verifiable
with the computationally-unbounded prover.
If we could modify the sum-check protocol for BQP problems
in such a way that
the honest prover's computational power is in quantum polynomial-time,
the open problem is solved. 
There are two results in this direction~\cite{AharonovGreen,Yupan}.
\item
Specific problems:
Several specific problems in BQP, such as the recursive Fourier sampling,
problems related to circuits in the second level of the Fourier hierarchy,
and calculating orders of solvable groups, are
classically verifiable~\cite{MattFourier,Tommaso,FH2,LeGall}.
\item
Rational prover: For any BQP problem,
it is possible to construct a rational proof system~\cite{AM} 
where a classical verifier sends a reward to the prover in such a way
that the prover who wants to maximize its profit has to send a correct 
solution to the verifier~\cite{rational,rationalTakeuchi}.
\end{itemize}

The simplest protocol in the first approach 
is so-called the posthoc verification~\cite{posthoc}.
(A detailed explanation of the posthoc verification is given in
Sec.~\ref{sec:preliminary}.)
In this protocol, any BQP problem can be verified in a non-interactive
way with a verifier who can do only single-qubit measurements:
the prover sends a quantum state to the verifier, and the verifier
measures each qubit of the state.
The prover can send each qubit of the state one by one,
and the verifier has only to measure each qubit sequentially,
i.e., the verifier does not need any quantum memory.
The idea of the posthoc verification is based on the
observations that the local Hamiltonian problem is 
QMA-complete~\cite{Kitaev,KKR},
BQP is in QMA with a trivial witness state (such as the all zero state), 
and the ground state of the local Hamiltonian
(i.e., the history state)
can be constructed in quantum polynomial-time if the corresponding
problem is in BQP~\cite{posthoc}.
Because the 2-local $XZ$-Hamiltonian problem
is QMA-complete~\cite{BL08,CM16}, the verifier
has only to measure randomly chosen two qubits in the computational
or the Hadamard basis.

A disadvantage of the posthoc protocol is, however,
that the verifier has to do the quantum measurements, and
the quantum channel from the prover to the verifier is required.
The Mahadev's breakthrough protocol~\cite{Mahadev}
removes them by using the cryptographic technique, but
the soundness becomes the
computational one, i.e., the protocol is an argument system.

In this paper, we show that if a trusted center is introduced,
an information-theoretically-sound
non-interactive 
verification of quantum computing is possible
for a classical verifier.
The trusted center sends random BB84 states to the prover,
and their classical descriptions to the verifier.
(Because the BB84 states are uniformly random, center's messages
are independent of the instance.)
Introducing a trusted center that distributes BB84 states
is somehow an artificial assumption, but it is not unrealistic
(for example, it is a foreseeable future
that the NIST distributes
BB84 states among quantum computing companies like Google, IBM, etc.), 
and the introduction of the trusted center gives us
huge advantages, namely, the classical verifier,
the non-interactiveness, and the information-theoretical
soundness.

More precisely, for each instance $x\in A$ of
any promise problem $A=(A_{yes},A_{no})$ in BQP, we consider the protocol
of Fig.~\ref{protocol:our}, and show
its completeness and soundness.
(A proof is given in Sec.~\ref{sec:proof}.)

\begin{theorem}
\label{theorem:newposthoc}
For any promise problem $A=(A_{yes},A_{no})$ in BQP,
the protocol of Fig.~\ref{protocol:our} satisfies both of the following
with $c$ and $s$ such that $c-s\ge1/poly(|x|)$:
\begin{itemize}
\item
If $x\in A_{yes}$, there exists a quantum polynomial-time prover such that 
the acceptance probability of the verifier is at least $c$.
\item
If $x\in A_{no}$,
the verifier's acceptance probability is at most $s$ for any
prover (even for computationally-unbounded prover).
\end{itemize}
\end{theorem}

\begin{figure}[h]
\rule[1ex]{\textwidth}{0.5pt}
\begin{itemize}
\item[1.]
The trusted center uniformly randomly chooses
$(h,m_1,...,m_N)\in\{0,1\}^{N+1}$,
where $N=poly(|x|)$. 
The trusted center sends $\bigotimes_{j=1}^N(H^h|m_j\rangle)$ 
to the prover.
The trusted center sends $(h,m_1,...,m_N)$ to the verifier.
Note that because $(h,m_1,...,m_N)$ is uniformly randomly chosen,
the messages from the trusted center to the prover and the verifier
are independent of the instance $x$.
\item[2.]
The prover does a POVM measurement, which can be done in 
quantum polynomial-time if the prover is honest, on the received state,
and sends the measurement
result (a $2N$-bit classical bit string) to the verifier.
\item[3.]
The verifier does a certain polynomial-time classical computation
to make the decision, accept/reject.
\end{itemize} 
\rule[1ex]{\textwidth}{0.5pt}
\caption{A high-level description of our verification protocol.}
\label{protocol:our}
\end{figure}

Our classical verification protocol can also be modified
to construct
a non-interactive statistical zero-knowledge proof system for QMA
with the trusted center (Fig.~\ref{protocol:highlevelzk}).
We show its completeness, soundness, and statistical zero-knowledge
property.
(A proof is given in Sec.~\ref{sec:proofzk}.)

\begin{theorem}
\label{theorem:zk}
For any promise problem $A=(A_{yes},A_{no})$ in QMA,
the protocol of Fig.~\ref{protocol:highlevelzk} 
satisfies all of the following
with $c$ and $s$ such that $c-s\ge1/poly(|x|)$:
\begin{itemize}
\item
If $x\in A_{yes}$, there exists a quantum polynomial-time prover 
(that receives a witness state of QMA as input) such that 
the acceptance probability of the verifier is at least $c$.
\item
If $x\in A_{no}$,
the verifier's acceptance probability is at most $s$ for any
prover (even for computationally-unbounded prover).
\item
It is statistical zero-knowledge.
\end{itemize}
\end{theorem}

\begin{figure}[h]
\rule[1ex]{\textwidth}{0.5pt}
\begin{itemize}
\item[1.]
The trusted center uniformly randomly chooses
$(h,m_1,...,m_N)\in\{0,1\}^{N+1}$,
where $N=poly(|x|)$. 
The trusted center sends $\bigotimes_{j=1}^N(H^h|m_j\rangle)$ 
to the prover.
The trusted center uniformly randomly chooses $(a,b)$ such 
that $1\le a<b\le N$.
The trusted center sends $(h,a,b,m_a,m_b)$ to the verifier.
\item[2.]
The prover does a POVM measurement, which can be done in 
quantum polynomial-time (given the witness state)
if the prover is honest, on the received state,
and sends the measurement
result (a $2N$-bit classical bit string) to the verifier.
\item[3.]
The verifier does a certain polynomial-time classical computation
to make the decision, accept/reject.
\end{itemize} 
\rule[1ex]{\textwidth}{0.5pt}
\caption{A high-level description of our zero-knowledge protocol.}
\label{protocol:highlevelzk}
\end{figure}

The idea is based on the recent elegant construction of
zero-knowledge systems for QMA in Ref.~\cite{BroadbentGrilo}.
In their construction, the prover sends the verifier
the one-time-padded ground state of the Hamiltonian 
that corresponds to the encoded version of the verification circuit,
and a classical commitment of the one-time-pad key.
After receiving a challenge from the verifier,
the prover opens only a small part of the ground state,
which is enough to measure energy but
not enough to get any information about the witness
due to the local simulatability~\cite{BroadbentGrilo,GSY}.
A zero-knowledge system for QMA was first constructed
in Ref.~\cite{BJSW}, and improvements have been obtained
recently~\cite{CVZ,Alagic,BroadbentGrilo}. 

The rest of the paper is organized as follows.
In the next section, we give preliminaries: we review the posthoc
verification protocol.
In Sec.~\ref{sec:proof}, we show Theorem~\ref{theorem:newposthoc}.
In Sec.~\ref{sec:proofzk}, we show Theorem~\ref{theorem:zk}.
Finally, we give discussions in Sec.~\ref{sec:discussion}.

\section{Preliminaries}
\label{sec:preliminary}
Let $A=(A_{yes},A_{no})$ be a promise problem in BQP. For any 
instance $x\in A$,
we can construct an $N$-qubit local Hamiltonian
\begin{eqnarray*}
H\equiv\sum_{i<j}
\frac{p_{i,j}}{2}\Big(\frac{I+s_{i,j} X_i\otimes X_j}{2} 
+\frac{I+s_{i,j} Z_i\otimes Z_j}{2} 
\Big)
\end{eqnarray*}
with $N=poly(|x|)$
such that if $x\in A_{yes}$ then the ground energy is less than $\alpha$,
and if $x\in A_{no}$ then the ground energy is larger than $\beta$
with $\beta-\alpha\ge1/poly(|x|)$.
Here,
$p_{i,j}>0$, $\sum_{i<j} p_{i,j}=1$, and $s_{i,j}\in\{+1,-1\}$.
The posthoc protocol~\cite{posthoc} runs as in Fig.~\ref{protocol:posthoc}.
Note that in this protocol, the verifier does not need any
quantum memory, because the verifier has only to measure each qubit
sequentially.
The verifier's acceptance probability is
$
p_{acc}=1-\mbox{Tr}(\rho H).
$
Therefore, if $x\in A_{yes}$, 
$p_{acc}\ge1-\alpha$ for an honest prover, and if $x\in A_{no}$,
$p_{acc}\le1-\beta$ for any (computationally-unbounded) prover.
The completeness-soundness gap is
$(1-\alpha)-(1-\beta)=\beta-\alpha\ge1/poly(|x|)$.
It is easy to see that this protocol can be done in the parallel way
to amplify the completeness-soundness gap.

\begin{figure}[h]
\rule[1ex]{\textwidth}{0.5pt}
\begin{itemize}
\item[1.]
The prover sends an $N$-qubit state $\rho$ to the verifier.
(If the prover is honest, it is the ground state of $H$.)
\item[2.]
The verifier uniformly randomly chooses $h\in\{0,1\}$.
\item[3.]
If $h=0$ ($h=1$), 
the verifier measures all qubits of $\rho$ in 
the computational (Hadamard) basis.
Let $m_j\in\{0,1\}$ be the measurement result on the $j$th qubit.
\item[4.]
The verifier samples $(i,j)$ with probability $p_{i,j}$.
\item[5.]
If $(-1)^{m_i}(-1)^{m_j}=-s_{i,j}$, the verifier accepts.
Otherwise, reject.
\end{itemize}
\rule[1ex]{\textwidth}{0.5pt}
\caption{The posthoc protocol~\cite{posthoc}.}
\label{protocol:posthoc}
\end{figure}

There is a remark: In the original posthoc protocol~\cite{posthoc},
the verifier first samples $(i,j)$ with probability $p_{i,j}$ and
measures $i$th and $j$th qubits.
The protocol explained in Fig.~\ref{protocol:posthoc} is slightly
modified from the original posthoc protocol in such a way that
the verifier's measurement is independent of the instance $x$.
Such a modification was already
done in Refs.~\cite{Alagic,BroadbentGrilo,Vidick_review},
and in fact the modification is crucial for our purpose, because the
trusted center's message should be independent of the instance $x$.

\section{Proof of Theorem~\ref{theorem:newposthoc}}
\label{sec:proof}
In this section, we show Theorem~\ref{theorem:newposthoc}.
Let us first
consider the protocol of Fig.~\ref{protocol:v1}
that we call the virtual protocol 1.
It is easy to see that the virtual protocol 1
has the same completeness and soundness as
those of the posthoc protocol (Fig.~\ref{protocol:posthoc}).
Next let us consider the protocol
of Fig.~\ref{protocol:v2}
that we call the virtual protocol 2.
The difference from the virtual protocol 1 is that the verifier first
measures halves of Bell pairs before sending other halves to the prover.
Because the verifier's measurement and the prover's measurement
commute with each other,
verifier's acceptance probability of the virtual protocol 2
is the same as that of the virtual protocol 1.
Finally, let us consider the protocol of Fig.~\ref{protocol:our2},
which is our final protocol.
The difference from the virtual protocol 2 is that the verifier's quantum
task
is done by the trusted center.
It is clear that the verifier's acceptance probability of this protocol
is the same as that of the virtual protocol 2.
In conclusion, our protocol (Fig.~\ref{protocol:our2})
has the same completeness and soundness
as those of the posthoc protocol (Fig.~\ref{protocol:posthoc}).

\begin{figure}[h]
\rule[1ex]{\textwidth}{0.5pt}
\begin{itemize}
\item[1.]
The verifier uniformly randomly
chooses $h\in\{0,1\}$.
The verifier generates $N$ Bell pairs,
and sends halves to the prover.
\item[2.]
If the prover is honest, the prover teleports the ground state of $H$
to the verifier by using the Bell pairs,
and sends the verifier the information
$(x,z)\in\{0,1\}^N\times\{0,1\}^N$ 
about the byproduct caused by the teleportation.
If the prover is malicious, the prover does any POVM 
measurement
$\{\Pi_{x,z}\}_{(x,z)\in\{0,1\}^N\times\{0,1\}^N}$ 
on the received states,
and sends the result
$(x,z)\in\{0,1\}^N\times\{0,1\}^N$ 
of the POVM measurement
to the verifier.
\item[3.]
If $h=0$ ($h=1$), the verifier measures all qubits of the teleported state
in the computational (Hadamard) basis.
Let $m_j\in\{0,1\}$ be the measurement result of the $j$th qubit.
\item[4.]
Let us define
$m_j'\equiv m_j\oplus (hz_j+(1-h)x_j)$ for $j=1,2,...,N$,
which are the measurement results that take into account the effects
of the teleportation byproducts.
The verifier samples $(i,j)$ with probability $p_{i,j}$.
The verifier accepts
if and only if $(-1)^{m_i'}(-1)^{m_j'}=-s_{i,j}$.
\end{itemize}
\rule[1ex]{\textwidth}{0.5pt}
\caption{The virtual protocol 1.}
\label{protocol:v1}
\end{figure}

\begin{figure}[h]
\rule[1ex]{\textwidth}{0.5pt}
\begin{itemize}
\item[1.]
The verifier uniformly randomly chooses $h\in\{0,1\}$.
The verifier generates $N$ Bell pairs.
\item[2.]
The verifier measures halves of the Bell pairs
in the computational (Hadamard) basis if $h=0$ ($h=1$).
Let $m_j\in\{0,1\}$ be the measurement result for the $j$th Bell pair.
The verifier sends unmeasured halves of the Bell pairs
to the prover.
\item[3.]
The prover does a POVM 
measurement
$\{\Pi_{x,z}\}_{(x,z)\in\{0,1\}^N\times\{0,1\}^N}$,
which corresponds to the teleportation of the ground state
when the prover is honest,
on the received states,
and sends the result
$(x,z)\in\{0,1\}^N\times\{0,1\}^N$ 
of the POVM measurement
to the verifier.
\item[4.]
The same as the step 4 of the virtual protocol 1.
\end{itemize}
\rule[1ex]{\textwidth}{0.5pt}
\caption{The virtual protocol 2.}
\label{protocol:v2}
\end{figure}

\begin{figure}[h]
\rule[1ex]{\textwidth}{0.5pt}
\begin{itemize}
\item[1.]
The trusted center uniformly randomly chooses
$(h,m_1,...,m_N)\in\{0,1\}^{N+1}$.
The trusted center sends $\bigotimes_{j=1}^N(H^h|m_j\rangle)$
to the prover. 
The trusted center sends $(h,m_1,...,m_N)$ to the verifier.
\item[2.]
The same as the steps 3 and 4 of the virtual protocol 2.
\end{itemize}
\rule[1ex]{\textwidth}{0.5pt}
\caption{Our protocol.}
\label{protocol:our2}
\end{figure}

\section{Proof of Theorem~\ref{theorem:zk}}
\label{sec:proofzk}
Our non-interactive statistical zero-knowledge
proof system for QMA with the trusted center is shown in
Fig.~\ref{protocol:zk}.
To show its completeness, soundness, and zero-knowledge property,
let us consider the protocol of Fig.~\ref{protocol:zk_v},
which we call the virtual zero-knowledge protocol.
It is easy to verify that protocols of
Fig.~\ref{protocol:zk} and Fig.~\ref{protocol:zk_v}
are the same.
The verifier's acceptance probability in 
the virtual zero-knowledge protocol is
$
p_{acc}
=1-\frac{1}{{N\choose 2}}\mbox{Tr}(\rho H),
$
and therefore the completeness-soundness gap is $1/poly(|x|)$.
The zero-knowledge property is also clear,
because
in the virtual zero-knowledge protocol,
what the verifier gets under the honest prover are
the uniformly randomly chosen $(h,a,b,x,z)$,
and the measurement results $(m_a,m_b)$ on the
$a$th and $b$th qubits of the one-time padded
history state in the base $h$ that is simulatable
in classical polynomial-time 
due to 
the local simulatability of the history state~\cite{BroadbentGrilo,GSY}.
In Fig.~\ref{protocol:simulator}, we show the simulator.
It is clear that the output of the simulator and verifier's view
are negligibly close.

\begin{figure}[h]
\rule[1ex]{\textwidth}{0.5pt}
\begin{itemize}
\item[1.]
The trusted center uniformly randomly chooses
$(h,m_1,...,m_N)\in\{0,1\}^{N+1}$.
The trusted center sends $\bigotimes_{j=1}^N(H^h|m_j\rangle)$
to the prover. 
The trusted center uniformly randomly chooses $(a,b)$ such that
$1\le a<b\le N$.
The trusted center sends $(h,a,b,m_a,m_b)$ to the verifier.
\item[2.]
The same as the step 3 of the virtual protocol 2.
\item[3.]
Let us define $m_j'\equiv m_j\oplus(hz_j+(1-h)x_j)$
for $j=a,b$.
The verifier samples $(i,j)$ with probability $p_{i,j}$.
If $a\neq i$, the verifier accepts.
If $b\neq j$, the verifier accepts.
If $a=i$ and $b=j$, 
the verifier accepts if and only if 
$(-1)^{m_a'}(-1)^{m_b'}=-s_{a,b}$.
\end{itemize}
\rule[1ex]{\textwidth}{0.5pt}
\caption{The zero-knowledge protocol.}
\label{protocol:zk}
\end{figure}

\begin{figure}[h]
\rule[1ex]{\textwidth}{0.5pt}
\begin{itemize}
\item[1.]
The trusted center uniformly randomly chooses
$h\in\{0,1\}$.
The trusted center generates $N$ Bell pairs, and sends halves to the prover.
\item[2.]
The prover does a POVM measurement, $\{\Pi_{x,z}\}_{(x,z)\in\{0,1\}^N
\times\{0,1\}^N}$, which corresponds to the teleportation of the
ground state when the prover is honest, on the received states.
\item[3.]
The trusted center measures the halves of the Bell pairs in the
computational (Hadamard) basis if
$h=0$ ($h=1$). Let $m_j\in\{0,1\}$ be the measurement result
for the $j$th Bell pair.
\item[4.]
The trusted center randomly chooses $(a,b)$ such that
$1\le a<b\le N$, and sends $(h,a,b,m_a,m_b)$ to the verifier.
\item[5.]
The prover sends the result $(x,z)\in\{0,1\}^N\times\{0,1\}^N$ of the 
POVM measurement to the verifier.
\item[6.]
The same as the step 3 of the zero-knowledge 
protocol~Fig.\ref{protocol:zk}.
\end{itemize}
\rule[1ex]{\textwidth}{0.5pt}
\caption{The virtual zero-knowledge protocol.}
\label{protocol:zk_v}
\end{figure}

\begin{figure}[h]
\rule[1ex]{\textwidth}{0.5pt}
\begin{itemize}
\item[1.]
The simulator uniformly randomly generates $h\in\{0,1\}$.
\item[2.]
The simulator uniformly randomly generates $(x,z)\in\{0,1\}^N\times\{0,1\}^N$.
\item[3.]
The simulator uniformly randomly generates $(a,b)$ such that
$1\le a<b\le N$.
\item[4.]
The simulator computes the classical description of 
$\rho_{a,b}\equiv{\rm Tr}_{a,b}(X^xZ^z|G\rangle\langle G|Z^zX^x)$,
where $|G\rangle$ is the ground state of the Hamiltonian.
\item[5.]
The simulator samples the measurement results
$(m_a,m_b)\in\{0,1\}^2$ in the basis $h$ on $\rho_{a,b}$.
\item[6.]
The simulator outputs $(h,a,b,m_a,m_b,x,z)$.
\end{itemize}
\rule[1ex]{\textwidth}{0.5pt}
\caption{The simulator.}
\label{protocol:simulator}
\end{figure}

\section{Discussion}
\label{sec:discussion}
In this paper, we have constructed
an information-theoretically-sound
non-interactive classical verification protocol
for quantum computing
with 
a trusted center.
The trusted center sends randomly chosen BB84 states to the prover,
and their classical descriptions to the verifier.

\if0
In the non-interactive protocols, 
there are three types
of models, the common random string (CRS) model, 
the secret parameter model, and the
preprocessing model. In the CRS model, the uniformly random
string is broadcasted by a trusted center.
In the secret parameter model, the message to the prover and that to
the verifier can be different. Finally, in the preprocessing
model, there is no trusted center, and the prover
and the verifier exchange messages (that are independent of the instance)
before 
the prover sending the witness (that depends on the instance) to the verifier.
Our protocol could be considered as a ``quantum secret parameter model".
It is easy to see that the same task can be done in the ``quantum
CRS model", where the trusted center distributes Bell pairs
among the prover and the verifier.
This, however, does not seem to be useful in a practical situations, because
the verifier has to keep the entanglement with the prover.
\fi

One might ask
whether the quantum message from the center to the prover can be
replaced with a classical one.
It will be impossible, because
if it was possible, then BQP is in AM, which is unlikely.
To see it, assume that the trusted center sends some random
classical messages to the prover. Then,
the message can be sent from the verifier instead of the center,
and it is a two-message AM protocol.

The combination of the trusted center considerd in this paper and
the Fitzsimons-Kashefi protocol~\cite{FK} also realizes the
information-theoretically-sound
classical verification of quantum computing, but in that case,
the protocol is interactive: polynomially-many rounds are
necessary between the prover and the verifier.
Furthermore, the messages sent from the trusted center do depend
on the instance.

The trusted center's task considered 
in this paper 
can be done by the ``remote state preparation" protocol.
In the remote state preparation,
the classical verifier can remotely prepare random quantum states
in the prover's place with only a classical communication
in such a way that only the verifier
knows which states are prepared.
It is well known that if the remote state preparation is possible,
a classical verification of quantum computing is possible.
It is open whether an information-theoretically-sound
remote state preparation is possible or not,
but
it was shown recently that
computationally-sound remote state preparations are possible under the 
LWE assumption~\cite{Vazirani,AndruVidick,MetgerVidick,Cojocaru}. 
If we combine these remote state preparation
protocols with our protocol, 
we would obtain a computationally-sound non-interactive classical 
verification protocol for quantum computing with
preprocessing.

\acknowledgements
TM is supported by MEXT Q-LEAP, JST PRESTO No.JPMJPR176A,
and the Grant-in-Aid for Young Scientists (B) No.JP17K12637 of JSPS.


\begin{thebibliography}{00}
\bibitem{Gottesman}
D. Gottesman, 2004.
\url{http://www.scottaaronson.com/blog/?p=284}
\bibitem{AharonovVazirani}
D. Aharonov and U. Vazirani,
Is quantum mechanics falsifiable? A computational perspective
on the foundations of quantum mechanics.
arXiv:1206.3686
\bibitem{Andru_review}
A. Gheorghiu, T. Kapourniotis, and E. Kashefi,
Verification of quantum computation: an overview of existing
approaches.
Theory of Computing Systems {\bf63}, 715-808 (2019);
arXiv:1709.06984




\bibitem{FK}
J. F. Fitzsimons and E. Kashefi,
Unconditionally verifiable blind computation.
Phys. Rev. A {\bf96}, 012303 (2017).

\bibitem{Aharonov}
D. Aharonov, M. Ben-Or, E. Eban, and U. Mahadev,
Interactive proofs for quantum computations.
arXiv:1704.04487

\bibitem{HM}
M. Hayashi and T. Morimae,
Verifiable measurement-only blind quantum computing
with stabilizer testing.
Phys. Rev. Lett. {\bf115}, 220502 (2015).

\bibitem{MNS}
T. Morimae, D. Nagaj, and N. Schuch,
Quantum proofs can be verified using only single-qubit
measurements.
Phys. Rev. A {\bf93}, 022326 (2016).

\bibitem{posthoc}
J. F. Fitzsimons, M. Hajdu{\v s}ek, and T. Morimae,
Post hoc verification of quantum computation.
Phys. Rev. Lett. {\bf120}, 040501 (2018).

\bibitem{Barz}
S. Barz, J. F. Fitzsimons, E. Kashefi, and P. Walther,
Experimental verification of quantum computation.
Nat. Phys. {\bf9}, 727 (2013).

\bibitem{Chiara}
C. Greganti, M. C. Roehsner, S. Barz, T. Morimae, and
P. Walther,
Demonstration of measurement-only blind quantum computing.
New J. Phys. {\bf18}, 013020 (2016).

\bibitem{Andru}
A. Gheorghiu, E. Kashefi, and P. Wallden,
Robustness and device independence of verifiable blind quantum
computing.
New J. Phys. {\bf17}, 083040 (2015).

\bibitem{TakeuchiMorimae}
Y. Takeuchi and T. Morimae,
Verification of many-qubit states.
Phys. Rev. X {\bf 8}, 021060 (2018).


\bibitem{MattMBQC}
M. McKague,
Interactive proofs for BQP via self-tested graph states.
Theory of Computing {\bf12}, 1 (2016).


\bibitem{Ji}
Z. Ji,
Classical verification of quantum proofs.
Proceedings of the 48th annual ACM symposium on Theory
of Computing (STOC 2016) p.885 (2016).

\bibitem{RUV}
B. W. Reichardt, F. Unger, and U. Vazirani,
Classical command of quantum systems.
Nature {\bf496}, 456 (2013).

\bibitem{Grilo}
A. B. Grilo,
A simple protocol for verifiable delegation of quantum computation
in one round.
46th International Colloquium on Automata, Languages, and Programming 
(ICALP 2019).

\bibitem{Coladangelo}
A. Coladangelo, A. B. Grilo, S. Jeffery, and T. Vidick,
Verifier-on-a-Leash: new schemes for verifiable delegated
quantum computation, with quasilinear resources.
arXiv:1708.07359; EUROCRYPT 2019.

\bibitem{Regev}
O. Regev,
On lattices, learning with errors, random linear codes, and
cryptography.
J. ACM, 56(6):34-134:40, 2009.

\bibitem{Mahadev}
U. Mahadev,
Classical verification of quantum computations.
IEEE 59th Annual Symposium on Foundations of Computer Science (FOCS), 
Paris, 2018, pp.259-267; arXiv:1804.01082

\bibitem{AndruVidick}
A. Gheorghiu and T. Vidick,
Computationally-secure and composable remote state preparation.
IEEE 60th Annual Symposium on Foundations of Computer Science (FOCS), Baltimore, MD, USA, 2019, pp. 1024-1033; arXiv:1904.06320

\bibitem{Alagic}
G. Alagic, A. M. Childs, A. B. Grilo, and S-H. Hung,
Non-interactive classical verification of quantum computation.
arXiv:1911.08101

\bibitem{Yamakawa}
N-H. Chia, K-M. Chung, and T. Yamakawa,
Classical verification of quantum computations with efficient verifier.
arXiv:1912.00990



\bibitem{AharonovGreen}
D. Aharonov and A. Green,
A quantum inspired proof of ${\rm P}^{\#{\rm P}}\subseteq{\rm IP}$.
arXiv:1710.09078
\bibitem{Yupan}
A. Green, Y. Liu, and G. Kindler,
Towards a quantum-inspired proof for IP=PSPACE.
arXiv:1912.11611

\bibitem{MattFourier}
M. McKague,
Interactive proofs with efficient quantum prover for
recursive Fourier sampling.
Chicago Journal of Theoretical Computer Science
{\bf6}, 1 (2012).

\bibitem{Tommaso}
T. F. Demarie, Y. Ouyang, and J. F. Fitzsimons,
Classical verification of quantum circuits containing few basis changes.
Phys. Rev. A {\bf97}, 042319 (2018).

\bibitem{FH2}
T. Morimae, Y. Takeuchi, and H. Nishimura,
Merlin-Arthur with efficient quantum Merlin and quantum supremacy
for the second level of the Fourier hierarchy.
Quantum {\bf2}, 106 (2018).
%arXiv:1711.10605

\bibitem{LeGall}
F. Le Gall, T. Morimae, H. Nishimura, and Y. Takeuchi,
Interactive proofs with polynomial-time quantum prover for
computing the order of solvable groups.
43rd International Symposium on Mathematical Foundations of Computer Science (MFCS 2018); arXiv:1805.03385

\bibitem{AM}
P. D. Azar and S. Micali,
Rational proofs.
Proceedings of the 44th symposium on Theory of Computing (STOC'12), 1017 (2012).

\bibitem{rational}
T. Morimae and H. Nishimura,
Rational proofs for quantum computing.
Quant. Inf. Comput. {\bf20}, 0181-0193 (2020);
arXiv:1804.08868

\bibitem{rationalTakeuchi}
Y. Takeuchi, T. Morimae, and S. Tani,
Sumcheck-based delegation of quantum computing to rational server.
arXiv:1911.04734

\bibitem{Kitaev}
A. Y. Kitaev, A. H. Shen, and M. N. Vyalyi,
Classical and Quantum Computation
(American Mathematical Society, Boston, MA, USA (2002))
\bibitem{KKR}
J. Kempe, A. Kitaev, and O. Regev,
The complexity of the local Hamiltonian problem.
SIAM Journal of Computing {\bf35}, pp.1070-1097 (2006).



\bibitem{BL08}
J. D. Biamonte and P. J. Love,
Realizable Hamiltonians for universal adiabatic quantum computers.
Physical Review A {\bf78}, 012352 (2008).
\bibitem{CM16}
T. Cubitt and A. Montanaro,
Complexity classification of local Hamiltonian problems.
SIAM Journal on Computing {\bf45}, 268-316 (2016).

\bibitem{BroadbentGrilo}
A. Broadbent and A. B. Grilo,
Zero-knowledge for QMA from locally simulatable proofs,
arXiv:1911.07782

\bibitem{GSY}
A. B. Grilo, W. Slofstra, and H. Yuen,
Perfect zero knowledge for quantum multi-prover interactive
proofs,
FOCS 2019; arXiv:1905.11280

\bibitem{BJSW}
A. Broadbent, Z. Ji, F. Song, and J. Watrous,
Zero-knowledge proof systems for QMA.
In 2016 IEEE 57th Annual Symposium on Foundations of Computer
Science (FOCS), pages 31-40. (2016);
SIAM J. Comput. 49(2), 245-283 (2020).

\bibitem{CVZ}
A. Coladangelo, T. Vidick, and T. Zhang,
Non-interactive zero-knowledge arguments for QMA, with
preprocessing.
arXiv:1911.07546

\bibitem{Vidick_review}
T. Vidick,
Verifying quantum computations at scale: a cryptographic leash on
quantum devices.
Bull. Amer. Math. Soc. {\bf57}, 39-76 (2020).


\bibitem{Vazirani}
Z. Brakerski, P. Christiano, U. Mahadev, U. Vazirani,
and T. Vidick,
A Cryptographic Test of Quantumness and Certifiable Randomness from a Single 
Quantum Device, IEEE 59th Annual Symposium on Foundations of Computer Science 
(FOCS), Paris, 2018, pp. 320-331;
arXiv:1804.00640
\bibitem{MetgerVidick}
T. Metger and T. Vidick,
Self-testing of a single quantum device under computational assumptions.
arXiv:2001.09161
\bibitem{Cojocaru}
A. Cojocaru, L. Colisson, E. Kashefi, and P. Wallden,
QFactory: classically-instructed remote secret qubits preparation.
ASIACRYPT 2019; arXiv:1904.06303

%\bibitem{GMR89}
%S. Goldwasser, S. Micali, and C. Rackoff,
%The knowledge complexity of interactive proof systems.
%SIAM J. Comput. {\bf18}, 186-208 (1989).

\end{thebibliography}
\end{document}